# Phenomenological viability of string and $M$-theory scenarios [*]

J.A. CASAS$^{\S\flat}$ A. IBARRA$^{\S}$ and C. MUÑOZ$^{\dagger\flat}$

$^{\S}$ *Instituto de Estructura de la Materia, CSIC, Serrano 123, 28006-Madrid, Spain*
$^{\dagger}$ *Departamento de Física Teórica C–XI, Univ. Autónoma de Madrid, 28049-Madrid, Spain*
$^{\flat}$ *Instituto de Física Teórica, C–XVI, Univ. Autónoma de Madrid, 28049-Madrid, Spain*

## Abstract

We analyze the constraints that a correct phenomenology and the absence of dangerous charge and color breaking (CCB) minima or unbounded from below (UFB) directions impose on the parameter space of different superstring and $M$-theory scenarios. Namely, we analyze in detail the case where supersymmetry (SUSY) breaking is driven by non-vanishing dilaton and/or moduli F-terms in weakly and strongly coupled strings, and the specific case where the mechanism of SUSY breaking is gaugino condensation with or without the participation of non-perturbative contributions to the Kähler potential. The results indicate that, only in some small windows of the parameter space all the previous tests are succesfully passed. We also discuss the impact of non-universality of the soft breaking terms on CCB/UFB bounds.

October 5, 1998

---

[*]Research supported in part by: CICYT of Spain, under contracts AEN95-0195 (JAC, AI) and AEN97-1678-E (CM); and the European Union, under contracts CHRX/CT92-0004 (JAC, AI) and ERBFMRX CT96-0090 (CM).

The purpose of this work is to examine the viability of different superstring and $M$-theory scenarios with respect to some essential phenomenological requirements as well as the desirable absence of dangerous charge and color breaking (CCB) minima or unbounded from below (UFB) directions in the effective potential [1].

At present, there have been constructed quite attractive weakly or strongly coupled string scenarios that have definite predictions on the soft supersymmetry (SUSY)-breaking terms, which is essential for phenomenological analyses. These constructions include: weakly coupled strings where SUSY breaking is driven by non-vanishing dilaton ($S$) and/or moduli ($T$) F-terms; scenarios where SUSY is broken by multiple gaugino condensation or by non-perturbative effects on the Kähler potential; and strongly coupled strings described by $M$-theory compactified on a Calabi-Yau (times a segment). The aim of this paper is to study all these constructions from the above-mentioned phenomenological criteria[1].

More precisely, we will impose the following phenomenological requirements: a correct (and not unnatural) electroweak breaking, a correct mass for the top quark, and values of the SUSY particles above the experimental lower bounds. It should be remarked here that the condition of a correct mass for the top is not trivial at all and, as we will see, imposes severe restrictions on the models. The reason is that it is not always possible to choose the boundary condition of the top Yukawa coupling, $\lambda_{top}$, so that the physical (pole) top mass is reproduced, because the renormalization group (RG) infrared fixed point of $\lambda_{top}$ puts an upper bound on the running top mass $m_{top}$, namely $m_{top} \lesssim 197 \sin\beta$ GeV, where $\tan\beta = \langle H_2 \rangle / \langle H_1 \rangle$. Concerning CCB/UFB bounds [8]-[13], [5], a complete analysis of all the potentially dangerous directions in the field space of the minimal supersymmetric standard model (MSSM) was carried out in [10, 11]. It was shown there that the corresponding constraints on the soft parameter space are in general very strong. The presence of these instabilities may imply either that the corresponding model is inconsistent or that it requires non-trivial cosmology to justify that the universe eventually fell in the phenomenologically realistic (but local) minimum [13]

Let us briefly review some of the basic ingredients required for our analysis. The general form of the soft SUSY-breaking Lagrangian in the context of the MSSM is given

---

[1]Some previous work in this direction has been done in the literature [2]-[7]. In particular, it was shown in [5] that if SUSY breaking is driven by the F-term associated to the dilaton, there is no choice of parameters for which all the phenomenological requirements are fulfilled and the realistic vacuum is the global minimum.

$$\mathcal{L}_{soft} = \frac{1}{2}\sum_{a=1}^{3} M_a \overline{\lambda}_a \lambda_a - \sum_i m_i^2 |\phi_i|^2 - (A_{ijk}W_{ijk} + B\mu H_1 H_2 + \text{h.c.}), \quad (1.1)$$

where $W_{ijk}$ are the usual terms of the Yukawa superpotential of the MSSM with $i=Q_L, u_L^c, d_L^c, L_L, e_L^c, H_1, H_2$, and $\phi_i$, $\lambda_a$ are the canonically normalized scalar and gaugino fields respectively. The above-mentioned string scenarios give definite predictions for scalar masses, gaugino masses and soft trilinear terms ($m_i$, $M_a$ and $A_{ijk}$ respectively in (1.1)). The value of the bilinear term $B$ is more model dependent and deserves some additional comments. Indeed, $B$ depends crucially on the particular mechanism which generates the associated (electroweak size) $\mu$ term in the superpotential ($\mu H_1 H_2$). Though there are interesting proposals for such a mechanism, there is no agreement about the most satisfactory one. In consequence we will leave $B$ as a free parameter in our analysis. On the other hand, as usual, we will fix the value of the $\mu$ parameter from the requirement of correct electroweak breaking, i.e. imposing the existence of a realistic minimum with the correct mass for the $Z$ boson.

Concerning the UFB directions (and corresponding constraints), they were analyzed in detail for the MSSM in [10], where it was found that there are three types of constraints, labelled as UFB-1, UFB-2, UFB-3. The UFB-3 direction, which involves the fields $\{H_2, \nu_{L_i}, e_{L_j}, e_{R_j}\}$ with $i \neq j$, yields the *strongest* bound among *all* the UFB and CCB constraints. For future convenience, let us briefly give the explicit form of this constraint. By simple analytical minimization of the relevant terms of the scalar potential it is possible to write the value of the $\{\nu_{L_i}, e_{L_j}, e_{R_j}\}$ fields in terms of the $H_2$ one. Then, for any value of $|H_2| < M_{string}$ satisfying

$$|H_2| > \sqrt{\frac{\mu^2}{4\lambda_{e_j}^2} + \frac{4m_{L_i}^2}{g'^2 + g_2^2}} - \frac{|\mu|}{2\lambda_{e_j}}, \quad (1.2)$$

the value of the potential along the UFB-3 direction is simply given by

$$V_{\text{UFB-3}} = (m_{H_2}^2 + m_{L_i}^2)|H_2|^2 + \frac{|\mu|}{\lambda_{e_j}}(m_{L_j}^2 + m_{e_j}^2 + m_{L_i}^2)|H_2| - \frac{2m_{L_i}^4}{g'^2 + g_2^2}. \quad (1.3)$$

Otherwise

$$V_{\text{UFB-3}} = m_{H_2}^2 |H_2|^2 + \frac{|\mu|}{\lambda_{e_j}}(m_{L_j}^2 + m_{e_j}^2)|H_2| + \frac{1}{8}(g'^2 + g_2^2)\left[|H_2|^2 + \frac{|\mu|}{\lambda_{e_j}}|H_2|\right]^2. \quad (1.4)$$

In (1.3,1.4) $\lambda_{e_j}$ is the leptonic Yukawa coupling of the $j$−generation and $m_{H_2}^2$ is the $H_2$ squared soft mass. Then, the UFB-3 condition reads

$$V_{\text{UFB-3}}(Q = \hat{Q}) > V_{\text{real min}}, \quad (1.5)$$

realistic minimum evaluated at the typical scale of SUSY masses, say $M_S$ (normally a good choice for $M_S$ is an average of the stop masses, for more details see [9, 14, 10]) and the renormalization scale $\hat{Q}$ is given by $\hat{Q} \sim \text{Max}(\lambda_{top}|H_2|, M_S)$. Notice from (1.3, 1.4) that the negative contribution to $V_{UFB-3}$ is essentially given by the $m_{H_2}^2$ term, which can be very sizeable in many instances. On the other hand, the positive contribution is dominated by the term $\propto 1/\lambda_{e_j}$, thus the larger $\lambda_{e_j}$ the more restrictive the constraint becomes. Consequently, the optimum choice of the $e$–type slepton is the third generation one, i.e. $e_j = $ stau.

Concerning the CCB constraints, let us mention that the "traditional" CCB bounds [8], when correctly evaluated (i.e. including the radiative corrections in a proper way), turn out to be extremely weak. However, the "improved" set of analytic constraints obtained in [10], which represent the necessary and sufficient conditions to avoid dangerous CCB minima, is much stronger. It is not possible to give here an account of the explicit form of the CCB constraints used in the present paper. This can be found in section 5 of [10], to which we refer the interested reader.

The paper is organized as follows. Sect. 2 is devoted to general models of weakly coupled strings without specifying the mechanism of SUSY breaking. We examine there several representative and well motivated scenarios. The results strongly suggest a most favorable model, which is discussed in one of the subsections. Sect. 3 is devoted to string scenarios where SUSY is broken by multiple gaugino condensation. The case where SUSY is broken thanks to non-perturbative corrections to the Kähler potential is examined in sect. 4, while $M$-theory scenarios are addressed in sect. 5. The conclusions are presented in sect. 6.

## 2 Weakly coupled string scenarios

In general, the soft SUSY-breaking terms [15] depend on the three functions which define a supergravity (SUGRA) theory, namely the Kähler potential $K(\phi_i, \phi_i^*)$, the gauge kinetic functions $f_a(\phi_i)$ and the superpotential $W(\phi_i)$. Here $\phi_i$ represent all the chiral matter fields. The soft breaking terms depend also on which is (are) the field(s) responsible for the SUSY breaking, i.e. the field(s) with non-vanishing F-term. These fields must belong to some hidden sector of the theory. In this section we will assume that the SUSY breaking is driven by the dilaton ($S$) F-term and/or the moduli ($T_j$) F-term [16, 17, 4]. They are fields that are always present in the effective SUGRA coming from weakly coupled strings and have gravitationally suppressed interactions with the observable fields. Thus, they are good candidates for the fields of the SUSY we will assume for simplicity that all of them get the same vacuum expectation value (VEV), which is the so-called overall modulus ($T$) approximation.

As is well known, the tree-level expression for $f_a$ in any four-dimensional (weakly coupled) superstring is $f_a = k_a S$, where $k_a$ is the Kac-Moody level of the gauge factor. Normally (level one case) one takes $k_3 = k_2 = \frac{3}{5} k_1 = 1$. Since a possible threshold $T$ dependence may appear due to the contribution of massive string states in loops, then in general

$$f_a = k_a S + \Delta f_a(T) , \qquad (2.1)$$

where $\Delta f_a(T)$ has been computed for simple orbifold compactifications of the heterotic string. The result has the form

$$\Delta f_a(T) = -\frac{1}{16\pi^2}(b'_a - k_a \delta_{GS}) \, log \, \eta^4(T) , \qquad (2.2)$$

where $\eta(T)$ is the Dedekind function, $\delta_{GS}$ is a group independent constant (typically $-\delta_{GS} = O(10)$) and

$$b'_a = b_a + 2\sum_i T_a(\phi_i)(1 + n_i) , \qquad (2.3)$$

where $b_a$ is the one-loop $\beta$-function coefficient of the $G_a$ gauge group coupling, $T_a(\phi_i)$ is the index of $G_a$ generators in the $\phi_i$ representation, and $n_i$ are the modular weights of the $\phi_i$ fields ($n_i = -1, -2, ..., -5$). It should be kept in mind that Re $f_a$ is essentially the "unified" gauge coupling constant at the string scale, $M_{string} \simeq 0.5 \times g_{string} \times 10^{18}$ GeV [18], where Re $g_{string}^{-2} = \text{Re} f_a \simeq 2$.

On the other hand, for orbifold compactifications the tree-level Kähler potential has in general the form

$$K = -\log(S + S^*) - 3\log(T + T^*) + \sum_i (T + T^*)^{n_i} \phi_i \phi_i^* , \qquad (2.4)$$

Assuming cancellation of the cosmological constant and neglecting phases, one obtains from (2.1, 2.4) the soft terms at the $M_{string}$ scale [4]

$$\begin{aligned} M_a &= \sqrt{3} m_{3/2} \left( \frac{k_a \text{Re} S}{\text{Re} f_a} \sin\theta \right. \\ &\quad + \left. \cos\theta \frac{(b'_a - k_a \delta_{GS})(T + T^*)\hat{G}_2(T, T^*)}{32\pi^3 \sqrt{3} \text{Re} f_a} \right) , \end{aligned} \qquad (2.5)$$

$$m_i^2 = m_{3/2}^2 (1 + n_i \cos^2\theta) , \qquad (2.6)$$

$$A_{ijk} = -\sqrt{3} m_{3/2} [\sin\theta + \cos\theta \, \omega_{ijk}(T, T^*)] , \qquad (2.7)$$

$$\omega_{ijk}(T,T^*) = \frac{1}{\sqrt{3}}\left(3 + n_i + n_j + n_k - (T+T^*)\frac{\partial \lambda_{ijk}/\partial T}{\lambda_{ijk}}\right), \tag{2.8}$$

$\hat{G}_2 = -4\pi(\partial \eta(T)/\partial T)(\eta(T))^{-1} - 2\pi/(T+T^*)$, $\lambda_{ijk}$ are the Yukawa couplings in the superpotential and $m_{3/2}^2 = e^K|W|^2$ is the gravitino mass. Finally, the goldstino angle $\theta$ parameterizes the relative contribution of $S$ and $T$ to the SUSY breaking.

One-loop corrections to the Kähler potential can be incorporated into the previous expressions for the soft terms, (2.5–2.7), by making the replacements [4]

$$S + S^* \longrightarrow Y = S + S^* - \frac{\delta_{GS}}{8\pi^2}log(T+T^*) \tag{2.9}$$

$$\cos\theta \longrightarrow \left(1 - \frac{\delta_{GS}}{24\pi^2 Y}\right)^{-1/2}\cos\theta \tag{2.10}$$

In the next subsections we analyze the phenomenological requirements as well as CCB/UFB bounds on the previous soft terms in particularly interesting and illustrative cases.

## 2.1  $n_i = -1$

We consider here a scenario where all the observable particles have modular weight $-1$. This is the case when the observable sector belongs to the untwisted sector of an orbifold compactification. The corresponding formulas for the soft breaking terms are also valid for a Calabi-Yau compactification in the large-$T$ limit.

This scenario has other interesting aspects. It guarantees universality of the soft scalar masses (see (2.6)) and allows to study the pure moduli-dominated limit, $\sin\theta = 0$. (Notice that for $n_i < -1$ the squared scalar masses (2.6) become negative in this limit, thus leading to the appearance of instabilities.) It is also interesting to note that for $n_i = -1$ the Yukawa couplings $\lambda_{ijk}$ are $T$-independent since a cubic operator of matter fields has exactly the appropriate modular weight $(-3)$. Consequently the expression for $A_{ijk}$ given by (2.7) becomes independent of $\lambda_{ijk}$ and universal. Ignoring the one-loop corrections encoded in (2.2, 2.9, 2.10), the expressions of $M, m, A$ become greatly simplified [4]

$$\begin{aligned} M &= \sqrt{3}m_{3/2}\sin\theta \\ m^2 &= m_{3/2}^2 \sin^2\theta, \\ A &= -\sqrt{3}m_{3/2}\sin\theta, \end{aligned} \tag{2.11}$$

of the value of $\theta$. This means that the results obtained in [5] for the dilaton-dominance limit [17, 4] (i.e. for the particular value $\sin\theta = 1$) are valid (at tree level) for any $\theta$. Since the dilaton-dominance limit was proved to be inconsistent with the UFB and CCB bounds, this must be also the case here for any $\theta$; and in fact it is. In Fig. 1a we have plotted the regions of the parameter space forbidden from the various CCB and UFB bounds, as well as from the condition of a correct top mass and and the experimental lower bounds on supersymmetric masses, for two fixed values of $m$ (viz. $m = 100, 500$ GeV). Notice from (2.5, 2.6, 2.7) that all the soft parameters (but $B$) can be written in terms of $m_{3/2}, \theta$ or, equivalently, in terms of $m, \theta$; thus the variables $\theta, B$ used in Fig. 1a to expand the parameter space at fixed $m$. The results, as expected, are independent[2] of $\theta$ and the whole parameter space becomes excluded. The same is true for any other (reasonable) value of $m$.

Once the one-loop corrections (2.2, 2.9, 2.10) are incorporated the results are not changed, except in the region of small $\sin\theta$ (i.e. near the moduli-dominance limit, where $M, A, m$ tend to vanish). For the sake of the clarity of the discussion, let us re-write the soft breaking terms from (2.5, 2.6, 2.7) (one-loop corrected) using a value $\mathrm{Re}T \simeq 1.2$, suggested by several gaugino condensation analyses [19]. Moreover this value is close to the $T$–duality self-dual point, which is what one would normally expect in a duality invariant theory. Then [4]

$$m^2 \simeq m_{3/2}^2 \left[\sin^2\theta - 10^{-3}\delta_{GS}\right] , \tag{2.12}$$

$$A = -\sqrt{3}m_{3/2}\sin\theta , \tag{2.13}$$

$$\begin{aligned} M_3 &\simeq 1.0\, \sqrt{3}m_{3/2}\left[\sin\theta - (3+\delta_{GS})\, 4.6\times 10^{-4}\, \cos\theta\right] ,\\ M_2 &\simeq 1.06\, \sqrt{3}m_{3/2}\left[\sin\theta - (-1+\delta_{GS})\, 4.6\times 10^{-4}\, \cos\theta\right] ,\\ M_1 &\simeq 1.18\, \sqrt{3}m_{3/2}\left[\sin\theta - (\frac{-33}{5}+\delta_{GS})\, 4.6\times 10^{-4}\, \cos\theta\right] . \end{aligned} \tag{2.14}$$

Some comments are in order here. The coefficients in front of the gaugino masses (2.14) are not the ones directly obtained from (2.5, 2.1, 2.2). Instead, it has been assumed that (for phenomenological requirements) the values of $\frac{k_a Y}{2\mathrm{Re}f_a}$ in (2.5) are the suitable ones to get effective unification at $M_{GUT} \simeq 3 \times 10^{16}$ GeV [3]. This is a reflect of the extra (unknown) threshold contributions modify the gauge couplings in the right way.

---
[2]Strictly speaking the results for $\sin\theta$ positive and negative are different, since they correspond to a (simultaneous) flip of the signs of $M$ and $A$. However, the results of the analysis are identical since they are unchanged under the transformation $M \to -M, A \to -A, B \to -B$. This is apparent from Fig. 1a.

[3]These are actually obtained from (2.1, 2.2) for other values of $\mathrm{Re}T$ and certain assignments of modular weights to the observable fields, see subsect. 2.3 below.

Near the moduli-dominance limit, $\sin\theta \to 0$, the tree-level contributions to $m, M_a, A$ (proportional to $m_{3/2}\sin\theta$) become very small, demanding a large value of $m_{3/2}$ in order to keep them $O(1\text{ TeV})$. Consequently, the one-loop corrections (proportional to $m_{3/2}\cos\theta$) become relatively more and more important, and the gaugino masses (2.14) become more non-universal. This lack of universality, together with that of the prefactors of in (2.14), turns out to be crucial to open (small) allowed windows in the parameter space near $\sin\theta = 0$, as it is illustrated in Fig. 1b for the typical case $\delta_{GS} = -5$. The reason is the following. As mentioned in the introduction, the UFB-3 bound is the most restricting of all the UFB and CCB bounds. The potential along the UFB-3 direction, $V_{\text{UFB}-3}$, is given by (1.3, 1.4). Clearly, the more negative $m^2_{H_2}$ and the smaller $m^2_{L_i}, m^2_{e_j}$, the more negative $V_{\text{UFB}-3}$, and thus the more restrictive the bound becomes. $m^2_{H_2}$ is driven negative by the stop contribution to its RGE, $Q\ dm^2_{H_2}/dQ = 6(\lambda_{top}/4\pi)^2(m^2_{\tilde{t}_L} + m^2_{\tilde{t}_R}) + ..$, hence the smaller $M^2_3 (and thus m^2_{\tilde{t}})$, the weaker the UFB-3 bound becomes. On the contrary, the RGE of $m^2_{L_i}, m^2_{e_j}$ is dominated by $M^2_2, M^2_1$. Hence, the greater $M^2_2, M^2_1$, the greater $m^2_{L_i}, m^2_{e_j}$ and the weaker the UFB-3 bound becomes. In other words, a departure of gaugino universality in the direction $M^2_2, M^2_1 > M^2_3$ helps to rescue regions in the parameter space from the UFB-3 bound. This is exactly what occurs in (2.14), thus the allowed windows in Fig. 1b.

It is worth noticing that the strict moduli-dominance limit becomes forbidden in the plots of Fig. 1b. As a matter of fact, for different values of $m$, this scenario is forbidden either from the condition of a correct top mass or from the experimental lower bounds on supersymmetric masses (not from CCB or UFB) unless $m > 2$ TeV (not represented in Fig. 1b). This value is in clear conflict with naturality bounds for electroweak breaking [21, 14, 22].

Incidentally, the previous arguments and results show that in the MSSM the assumption of non-universal gaugino masses must be very helpful to avoid the otherwise strong UFB constraints. This fact has not been previously noted in the literature.

## 2.2 $n_i = -2$

The scenario where all the observable fields have modular weight $-2$ is interesting because it represents a typical case where all these fields belong to the same twisted sector, thus having universal soft masses, see (2.6). It corresponds in particular (but not only) to the $Z_3$ orbifold case, which has been extensively treated in the literature. A first observation is that the allowed range for $\theta$ is restricted from the requirement of

$$\cos^2 \theta \leq \frac{1}{2} , \qquad (2.15)$$

which in particular discards the moduli-dominated limit. It is also worth noticing that the Yukawa couplings, $\lambda_{ijk}$, are in this case non-trivial functions of $T$ [23]. Apparently, this is a serious shortcoming for the calculation of $A_{ijk}$ (see (2.7), (2.8)), and thus the evaluation of the CCB and UFB bounds[4]. However, the top Yukawa coupling represents a remarkable (and fortunate) exception. The reason is the following. The twisted Yukawa couplings are in general given by a series of terms, all of them suppressed by $e^{-c_i T}$ factors. Only when the involved fields belong to the same fixed point (or fixed torus) the first term in the series is $O(1)$ and independent of $T$, otherwise the coupling is suppressed. Consequently, the top Yukawa coupling, being $O(1)$, must be of this type. So, for this particular case we can ignore the $\partial \lambda / \partial T$ factors in (2.8, 2.7), thus getting $\omega = -\sqrt{3}$, $|A| = -\sqrt{3} m_{3/2}(\sin \theta - \sqrt{3} \cos \theta)$. The nice thing is that the top $A$-coupling is the only relevant one for the UFB-3 constraint, which is by far the strongest CCB/UFB constraint. (Notice that no $A$-coupling appears explicitely in (1.3, 1.4), though the top $A$-coupling participates indirectly through the RGE of the various parameters involved.) Thus, we can safely evaluate the UFB-3 constraint in the twisted case.

The numerical results are shown in Fig. 2. We have just plotted the allowed region of $\theta$ (see (2.15)) in the $[0, \pi]$ range. As discussed above, the results in the $[\pi, 2\pi]$ range are equivalent due to the invariance under the transformation $M \to -M, A \to -A, B \to -B$ (see footnote 2). Apart from a very narrow band close to the left limit of $\theta$ (difficult to appreciate from the figure) the whole parameter space becomes forbidden, as in the $n_i = -1$ case. Actually, the allowed narrow band corresponds to very large values of $m_{3/2}$ (see (2.6)), and thus of $M$ and $A$, which makes this small region unacceptable from the point of view of naturalness of the electroweak breaking.

The previous analysis relies on the tree-level expressions for $M, m, A$ (see first term of (2.5) and (2.6, 2.7)), which in particular imply gaugino universality at the string scale. However, as has been mentioned in subsect. 2.1, gauge coupling unification at the GUT scale requires the factors $\frac{k_a Y}{2 \mathrm{Re} f_a}$ to be (slightly) non-universal, thus leading to gaugino non-universality in (2.5). Numerically these factors should be as given in (2.14), so it is reasonable to assume that at the string scale

$$M_3 : M_2 : M_1 = 1.0 : 1.06 : 1.18 . \qquad (2.16)$$

---

[4]Actually, the Yukawa couplings for all the abelian orbifolds have been conputed in [24]. There are tens of them, and their values depend on the particular assignment of the fields to the orbifold fixed points or fixed tori. Hence, a complete analysis is possible but extremely cumbersome.

direction to rescue regions from the UFB-3 bound. In fact, once these corrections are included the allowed band in Fig. 2 becomes somewhat enhanced, becoming acceptable for naturalness requirements.

To summarize, the $n_i = -2$ case is "more allowed" than the $n_i = -1$ one, but still the acceptable windows are quite small.

## 2.3 ILR model

This is a model proposed by Ibañez, Lüst and Ross [25] of assignment of modular weights to observable fields, which has the nice feature of providing appropriately large string threshold corrections to fit the joining of gauge coupling constants at a scale $\simeq 10^{16}$ GeV [16, 25]. This assignment is given by

$$n_{Q_L} = n_{d_L^c} = -1 \;, \;\; n_{u_L^c} = -2 \;, \;\; n_{L_L} = n_{e_L^c} = -3 \;, \;\; n_{H_1} + n_{H_2} = -5, -4 \;. \quad (2.17)$$

The above values together with a Re$T \simeq 16$ lead to good agreement for $\sin^2 \theta_W$ and $\alpha_3$. This scenario is also interesting because it provides us with an explicit model with non-universal modular weights, and thus (in general) non-universal scalar masses. It should be noticed that this lack of universality is not dangerous for flavour changing neutral currents, as it is family independent. In this case, the allowed range of $\theta$ in order to avoid negative squared scalar masses (in particular, slepton masses) is

$$\cos^2 \theta \leq \frac{1}{3} \;. \quad (2.18)$$

As mentioned above, the results are equivalent when shifting $\theta \to \theta + \pi$. Since the modular weights of (2.17) are in general different than $-1$, the couplings are generically of the twisted type. This imposes the limitations discussed in the previous subsection about the analysis of the CCB bounds. However, just as in the $n_i = -2$ case, the UFB-3 bound can be safely analyzed, and, as we will see, that is all we need. The assignment given in (2.17) gives a certain freedom for the choice of the modular weights of the two Higgs doublets. We have analyzed the following cases

$$\begin{aligned} n_{H_1} &= -2, \quad n_{H_2} = -3 \\ n_{H_1} &= -3, \quad n_{H_2} = -2 \\ n_{H_1} &= -2, \quad n_{H_2} = -2 \;. \end{aligned} \quad (2.19)$$

The numerical results for the first one are shown in Fig. 3. The whole parameter-space is excluded, either by the condition of a correct top mass or by the UFB-3 bound, in all the cases. The results for the other two cases are completely analogous.

The previous results suggest how to construct a model which is "optimized" from the CCB and UFB constraints. Namely, it has been explained in subsect. 2.1 that the larger the stop mass, the more negative is $m_{H_2}^2$, due to its RG running. In consequence, $V_{\text{UFB}-3}$ becomes more negative and the UFB-3 bound stronger (see (1.3, 1.4 1.5)). Analogously, the smaller $m_{L_i}^2, m_{e_j}^2$, the more negative $V_{\text{UFB}-3}$, and thus the stronger the UFB-3 bound becomes. The most favorable case thus corresponds to small squark masses and large slepton masses. Hence, we consider the following family-independent assignment

$$n_{Q_L} = n_{d_L^c} = n_{u_L^c} = -2 \quad , \quad n_{L_L} = n_{e_L^c} = -1 \quad . \tag{2.20}$$

Of course we could optimize further the model by considering more negative modular weights for the squarks, e.g. $n = -3, -4$. But this restricts the allowed range for $\theta$ and, on the other hand, corresponds to much less frequent cases in explicit string constructions. Concerning the Higgs sector we consider the following cases

$$\begin{aligned} n_{H_1} &= -1, \quad n_{H_2} = -1 \\ n_{H_1} &= -1, \quad n_{H_2} = -2 \quad . \end{aligned} \tag{2.21}$$

In this scenario the allowed range for $\theta$ is as in (2.15). Again, the results are equivalent when shifting $\theta \to \theta + \pi$. The numerical results for the first case in (2.21) are shown in Fig. 4a. Clearly, the allowed windows are now wider than in the previously analyzed scenarios. So, certainly from the CCB and UFB point of view the assignment of (2.20) is preferred. Nevertheless, even in this case, the allowed region is not very large. The results are analogous for the second choice in (2.21).

As was discussed in subsect. 2.2 one may prefer to use non-universal gaugino masses at the string scale ($M_3 : M_2 : M_1 = 1.0 : 1.06 : 1.18$) from gauge unification arguments. Then the allowed regions become somewhat enhanced, as shown in Fig. 4b.

Let us note that a similar strategy can be followed in the MSSM to open windows in the parameter space from UFB constraints by using non-universal scalar masses (this can be used in addition to the gaugino non-universality discussed at the end of sect. 2.1).

## 3 SUSY breakdown by gaugino condensation

In the previous section there was no explicit reference to the dynamical origin of SUSY breaking. In this section we consider the case where gaugino condensation in the hidden sector is the actual mechanism for the breakdown of SUSY, which is in fact the most can be found in [26] and references therein.) For our purposes it is enough to say that for a hidden sector gauge group $G = \prod_b G_b$, gaugino condensation induces a non–perturbative superpotential

$$W^{np} = \sum_b d_b \frac{e^{-3k_b S/2\beta_b}}{[\eta(T)]^{6+(3k_b \delta_{GS}/8\pi^2 \beta_b)}} \quad , \tag{3.22}$$

where $\beta_b$ are the corresponding beta functions and $d_b$ are constants. One finds that if the Kähler potential is given by the perturbative expression (2.4, 2.9) and the hidden sector contains a single gaugino condensate, then the usual SUGRA potential has no minima in $S$, $T$, presenting a run-away behaviour. Hence, one is led to consider at least two condensates (the so-called racetrack models [27, 26]). Then, it is easy to get realistic models with SUSY breaking and sensible values of $S$, $T$ and $m_{3/2}^2 = e^K |W|^2$ at the minimum [26]. Incidentally, the SUSY breakdown is mainly produced along the $F_T$ direction (moduli dominance). On the other hand, the scalar potential is quite involved, so a numerical analysis is in general necessary. Fortunately, the results for the soft terms admit quite simple and useful parameterizations describing them very well (within 1% of accuracy). Next, we give these parameterizations for the common $k_3 = k_2 = \frac{3}{5}k_1 = 1$ case [2]

$$m_i^2 = m_{3/2}^2 \left[ 1 + n_i (0.078 + 0.65 \times 10^{-4} \, \delta_{GS}) \right] \quad , \tag{3.23}$$

$$M_a = \frac{1}{4\pi \mathrm{Re} f_a} \, m_{3/2} \, (-0.006 \, \delta_{GS} + 0.0029) \quad , \tag{3.24}$$

$$A_{ijk} = (0.28 + 1.15 \times 10^{-4} \delta_{GS}) (3 + n_i + n_j + n_k) \quad , \tag{3.25}$$

where $n_i$ are the corresponding modular weights. The parameterization for $M_a$ is strictly valid only for the $Z_3$ and $Z_7$ orbifolds, although for the remaining cases it also holds with acceptable accuracy for our purposes. In the expression for $A$, a contribution proportional to $\frac{\partial \lambda_{ijk}/\partial T}{\lambda_{ijk}}$, where $\lambda_{ijk}$ is the corresponding Yukawa coupling, has been neglected. This is justified for the top, as was discussed in subsect. 2.2. It is worth noticing that (3.23, 3.24) lead to a hierarchy $M \ll m$, which requires large values of $m$ in order to avoid the experimental bounds on gauginos. This hierarchy is softened for large values of $\delta_{GS}$. We have taken $\delta_{GS} = -90$ as a typical large value for the numerical analysis. On the other hand, the $\mu$–parameter is fixed, as usual, by the electroweak breaking condition, while the value of $B$ is left as a free parameter. Consequently, the scenario contains just two independent parameters, $\{m_{3/2}, B\}$, or alternatively $\{m, B/m\}$. The restrictive power of the phenomenological conditions and the CCB/UFB any region of the relevant parameter space. However, $m$ is forced to be $\gtrsim 2$ TeV in order to avoid the experimental bounds on gauginos. This is a notorious shortcoming of the model, since it is hardly compatible with a natural electroweak breaking [2]

## 4  SUSY breakdown by non-perturbative Kähler potentials

As has been mentioned in sect. 3, gaugino condensation is a most promising mechanism for SUSY breaking. However, its implementation in the perturbative string constructions requires the presence of more than one condensate. Furthermore, it is difficult in such a scenario to get the minimum of the potential at $V = 0$, as it would be desirable for the cosmological constant problem. These shortcomings may be solved if the Kähler potential, $K$, receives sizeable non-perturbative corrections, a point of view that has been advocated in [28]. Actually, stringy non-perturbative effects are expected to be important, even in weakly coupled strings. On the other hand, arguments based on holomorphy show that these effects can only be large for the Kähler potential (not for the superpotential and the gauge kinetic functions). As mentioned above, these corrections are in principle capable to stabilize the dilaton potential at $V = 0$ with just one condensate. Then, even without specifying the explicit form of $K$, one gets non-trivial predictions for the soft terms [29][5].

More precisely, assuming that the non-perturbative superpotential is dominated by a single gaugino condensate, $W \sim e^{-aS}$, with $a \simeq 18$ in order to get a correct scale of SUSY breaking, the relevant formulas for the scalar and gaugino masses, $m, M$, and for the coefficient of the trilinear scalar terms, $A$, are given by[6]

$$
\begin{aligned}
m^2 &= m_{3/2}^2 \ , \\
|M| &= m_{3/2} \frac{3g^2}{|K' - 2a|} \ , \\
|A| &= |M| \left| \frac{K'}{g^2} \right| \ ,
\end{aligned}
\tag{4.26}
$$

where $K' \equiv \partial K / \partial (\mathrm{Re} S)$ should be considered as an independent (unknown) parameter. It is interesting to mention that the SUSY breaking is mainly produced along the dilaton direction, thus the universality of the soft terms in (4.26).

---

[5]It is interesting however to mention that explicit forms of $K_{non-pert}$, justified by theoretical arguments, are actually able to implement the previous desired conditions [30, 29, 31].

[6]Notice that the expression for $A$ in eq.(35) of [29] has the numerator and the denominator switched round due to a misprint.

logical requirements and CCB/UFB bounds. As usual, the value of $\mu$ can be extracted from the electroweak breaking condition and the value of $B$ is considered as a free parameter. Consequently, for a fixed value of $m_{3/2}$ (or, equivalently, of $m$) we are left with two independent parameters for the analysis, $B$ and $K'$. The numerical results are shown in Fig. 6 for two typical values of $m$. The range of values for $K'$ in the figure, namely $15 \leq K' \leq 57$, has been chosen in order to avoid a large hierarchy $M \ll m$, undesirable for naturalness reasons, as discussed in sect. 3. Notice also that the signs of $M$ and $A$ in (4.26) are undefined. Fig. 6 shows the results just for positive $A$ and $M$. Since we are exploring negative as well as positive values of $B$, and the results are unchanged under the transformation $M \to -M, A \to -A, B \to -B$, it is clear that only the relative sign of $M$ and $A$ is relevant. Hence, the case when $M$ and $A$ are both negative is equivalent to the one shown in the figure. Finally, the case where $A$ and $M$ have opposite signs leads to very similar results. This comes from the fact that $M$ is small in most of the parameter space, so the results are in practice quite insensitive to the value and sign of $M$.

Impresively enough, the whole parameter space becomes forbidden, essentially due to the condition of a correct mass for the top. The regions that survive this condition are excluded by UFB bounds and/or CCB bounds and/or experimental limits on supersymmetric masses, as shown in the figure.

## 5  $M$-theory scenarios

Recently Hořava and Witten proposed that the strong coupling limit of $E_8 \times E_8$ heterotic string theory can be described by 11-dimensional SUGRA on a manifold with boundary where the two $E_8$ gauge multiplets are restricted to the two 10-dimensional boundaries respectively ($M$-theory) [32].

Some phenomenological implications of the strong-coupling limit of $E_8 \times E_8$ heterotic string theory have been studied by compactifying the 11-dimensional $M$-theory on a Calabi-Yau manifold times the eleventh segment [33]. The resulting 4-dimensional effective theory can reconcile the observed Planck scale $M_P = 1/\sqrt{8\pi G_N} \approx 2.4 \times 10^{18}$ GeV with the phenomenologically favored GUT scale $M_{GUT} \approx 3 \times 10^{16}$ GeV in a natural manner, providing an attractive framework for the unification of couplings [33, 34]. In this framework, one could get the correct values of $\alpha_{GUT}$ and $M_P$, which was problematic in the weakly coupled heterotic string theory. An additional phenomenological virtue of the $M$-theory limit is that the strong CP problem may be satisfactory solved by the axion mechanism [34, 35]. These phenomenological virtues have motivated many of effective SUGRA model is given by [34, 35, 36, 37]

$$K = -\ln(S + \bar{S}) - 3\ln(T + \bar{T}) + \left(\frac{3}{T + \bar{T}} + \frac{\alpha}{S + \bar{S}}\right)|\phi|^2 ,$$
$$f = S + \alpha T ,$$
$$W = \lambda_{ijk}\phi^i\phi^j\phi^k , \qquad (5.27)$$

where the Yukawa couplings $\lambda_{ijk}$ are constants and $\alpha$ is an integer coefficient characterizing the Calabi-Yau space.

Assuming that SUSY is spontaneously broken by the auxiliary components of the bulk moduli superfields in the model ($S$ and $T$) one obtains from the above formulas the following soft terms at a scale $\simeq 10^{16}$ GeV [38]

$$M = \frac{\sqrt{3}m_{3/2}}{1+\epsilon}\left(\sin\theta + \frac{\epsilon}{\sqrt{3}}\cos\theta\right) ,$$
$$m^2 = m_{3/2}^2 - \frac{3m_{3/2}^2}{(3+\epsilon)^2}\left\{\epsilon(6+\epsilon)\sin^2\theta + (3+2\epsilon)\cos^2\theta - 2\sqrt{3}\epsilon\sin\theta\cos\theta\right\} ,$$
$$A = -\frac{\sqrt{3}m_{3/2}}{3+\epsilon}\left\{(3-2\epsilon)\sin\theta + \sqrt{3}\epsilon\cos\theta\right\} , \qquad (5.28)$$

where

$$\epsilon = \frac{\alpha(T + \bar{T})}{S + \bar{S}} . \qquad (5.29)$$

Notice that the structure of these soft terms is qualitatively different than those of the weakly coupled heterotic string case (2.11) which can be recovered from (5.28) by taking the limit $\epsilon \to 0$. Whereas there are only two free parameters in (2.11), viz. $m_{3/2}$ and $\theta$, the $M$-theory result (5.28) is more involved due to the additional dependence on $\epsilon$. Nevertheless we can simplify the analysis, as discussed in [38], by taking into account that the value of $\epsilon$ must be of order one, to be precise $0 < \epsilon \simeq 1$, in order to produce the known values of $\alpha_{GUT}$. Several comments are in order [38]. First of all, some ranges of $\theta$ are forbidden since they lead to by negative scalar squared-masses. The smaller the value of $\epsilon$, the smaller the forbidden range becomes. In the weakly coupled heterotic string limit, there is no a priori forbidden range for $\theta$, since the squared scalar masses are always positive (see (2.11)). On the other hand, it is worth noticing that scalar masses are always smaller than gaugino masses in this $M$-theory scenario. As discussed in subsect. 2.1, in the weakly coupled heterotic string the limit $\sin\theta \to 0$ is problematic since all $M$, $A$, $m$ vanish. One then had to include the string one-loop corrections to the Kähler potential and gauge kinetic functions which modified the boundary conditions (2.11). This problem is not present in the $M$-theory limit since gaugino masses are always different from zero.

analysis, the value of $\mu$ can be extracted from the electroweak breaking condition and the value of $B$ is considered as a free parameter. Consequently, for a fixed value of $m_{3/2}$ (or, equivalently, of $m$) we are left with three independent parameters $B$, $\theta$ and $\epsilon$. In Fig. 7 the restrictive power of the top quark mass condition and the CCB/UFB constraints is shown for two representative values of $\epsilon$. The results are equivalent when shifting $\theta \to \theta + \pi$ since they correspond to a (simultaneous) flip of the signs of $M$ and $A$ (see footnote 2). At the end of the day the whole parameter space becomes forbidden. It is worth noticing that the smaller the value of $\epsilon$ the smaller the dependence on $\theta$ of the results becomes, the reason being that, as mentioned above, in this limit the weakly coupled heterotic string is recovered.

## 6  Conclusions

We have analyzed the constraints that a correct phenomenology (i.e. a correct electroweak breaking, a correct mass for the top and supersymmetric masses above the experimental bounds) and the absence of dangerous CCB or UFB directions impose on the parameter space of different well-motivated superstring and $M$-theory scenarios.

Concerning weakly coupled strings, we have studied in detail the case where SUSY breaking is driven by non-vanishing dilaton and/or moduli F-terms. The results depend on the modular weights of the matter fields, so we have analyzed several representative and physically relevant cases. Also, we have studied the scenarios where SUSY is broken by multiple gaugino condensation or by non-perturbative effects on the Kähler potential. Concerning strongly coupled strings, we have studied the viability of the soft terms emerging from the Hořava-Witten $M$-theory compactified on a Calabi-Yau (times a segment).

The results indicate that, excepting some small windows, all the regions of the parameter space of the analyzed scenarios conflict some phenomenological constraints (particularly the possibility to get a correct mass for the top or experimental bounds on SUSY masses), and/or CCB, UFB constraints (particularly the last ones). However, it must be kept in mind that the regions excluded from CCB/UFB constraints could in principle be cosmologically rescued if the early Universe "decided" (e.g. for thermal reasons) to fall into the realistic minimum [13]. In that case we could be living in a local (accidental) minimum, provided its lifetime is long enough, a certainly funny situation.

On the other hand, we have noticed how the lack of universality of the soft terms affects the restrictive power of CCB and (very specially) UFB bounds. In this sense, the slight non-universality of gaugino masses predicted in some string scenarios, is able universality of scalar masses (with no conflict with FCNC), as allowed in some string scenarios, is able to open some windows in the parameter space. Following this line, we have constructed an optimized model, which has appreciably large areas allowed by all the constraints (see subsect. 2.4). The same kind of strategy, combining non-universality of gaugino and scalar masses, can be applied to the MSSM to open large windows in its parameter space from CCB/UFB bounds (see discussion in subsections 2.1, 2.4).

As this manuscript was prepared, ref.[39] appeared. That paper discusses some of the issues presented in this work, particularly the scenario of sect. 5

**Acknowledgments**

The work of A. Ibarra has been supported in part by a Comunidad de Madrid grant.

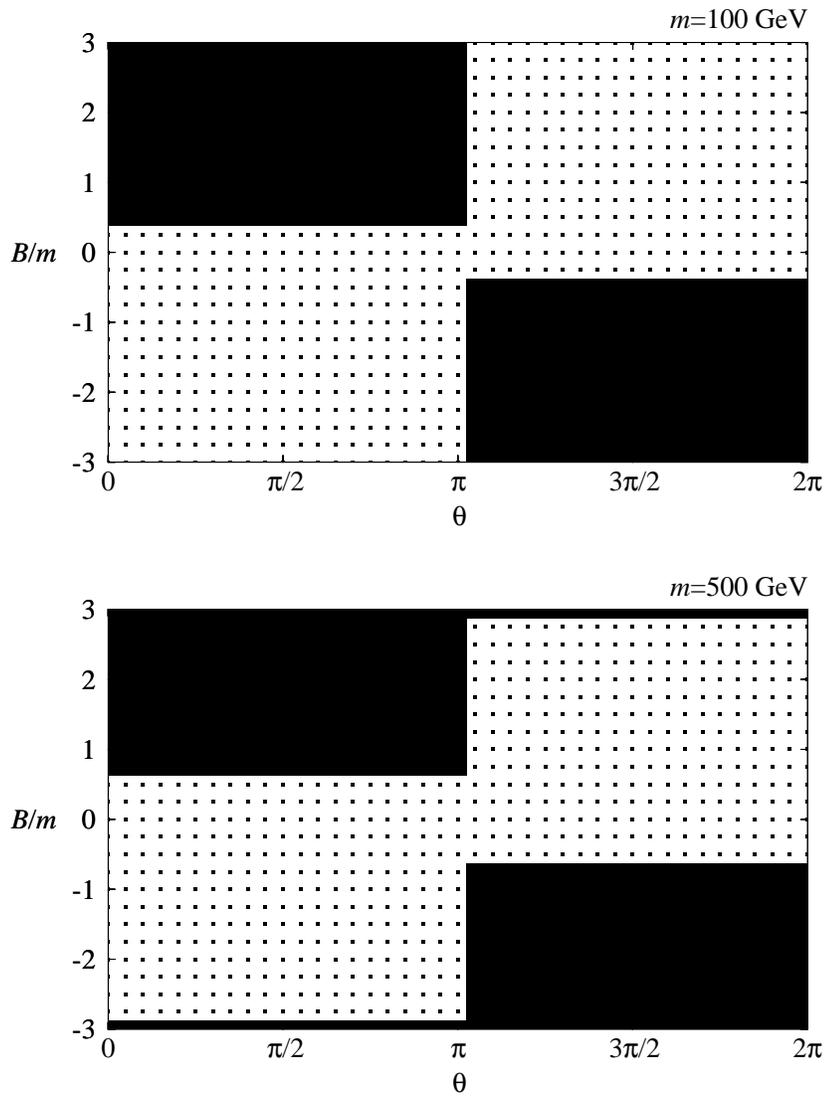

Fig.1a
Excluded regions of the parameter space of a weakly coupled string scenario where all the matter fields have modular weight $n_i = -1$ (see subsect. 2.1), for two different values of the soft scalar mass, $m$. The black region is excluded because it is not possible to reproduce the experimental mass of the top. The small squares indicate regions excluded by UFB constraints.

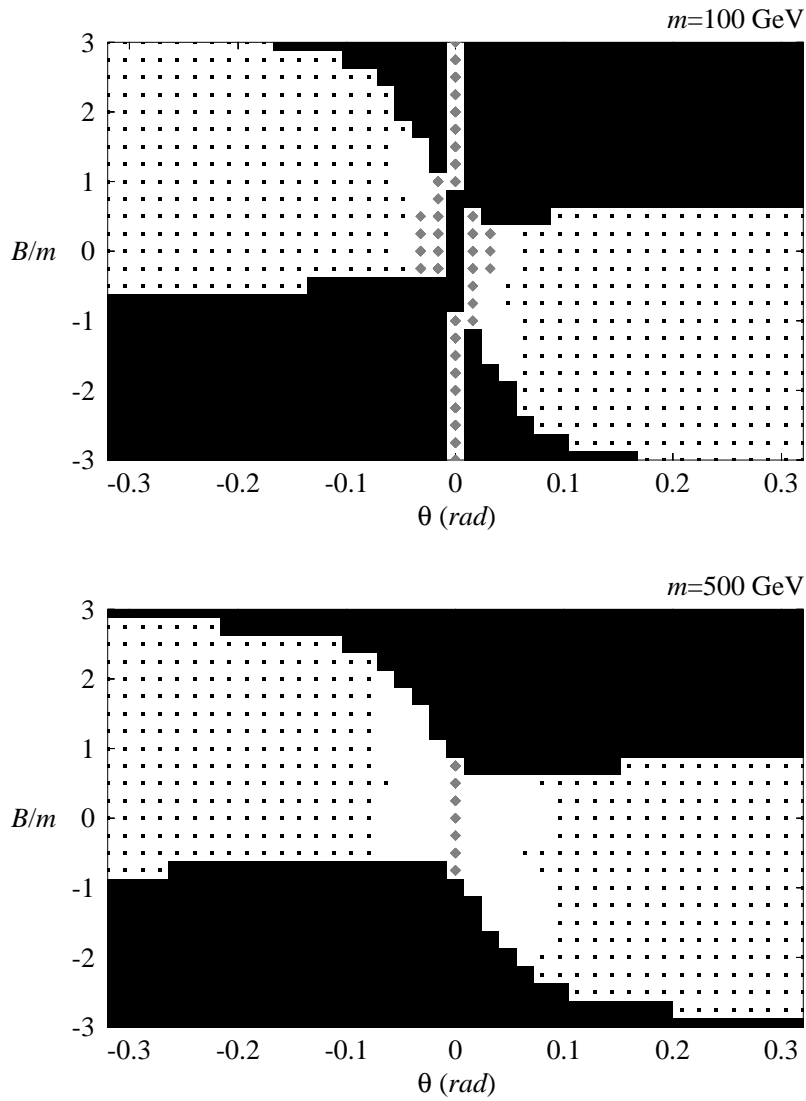

Fig.1b
The same as Fig.1a, but near the moduli-dominance limit ($\theta$ close to 0) and including one-loop corrections to the soft terms. The diamonds correspond to regions excluded by the experimental lower bounds on supersymmetric particles.

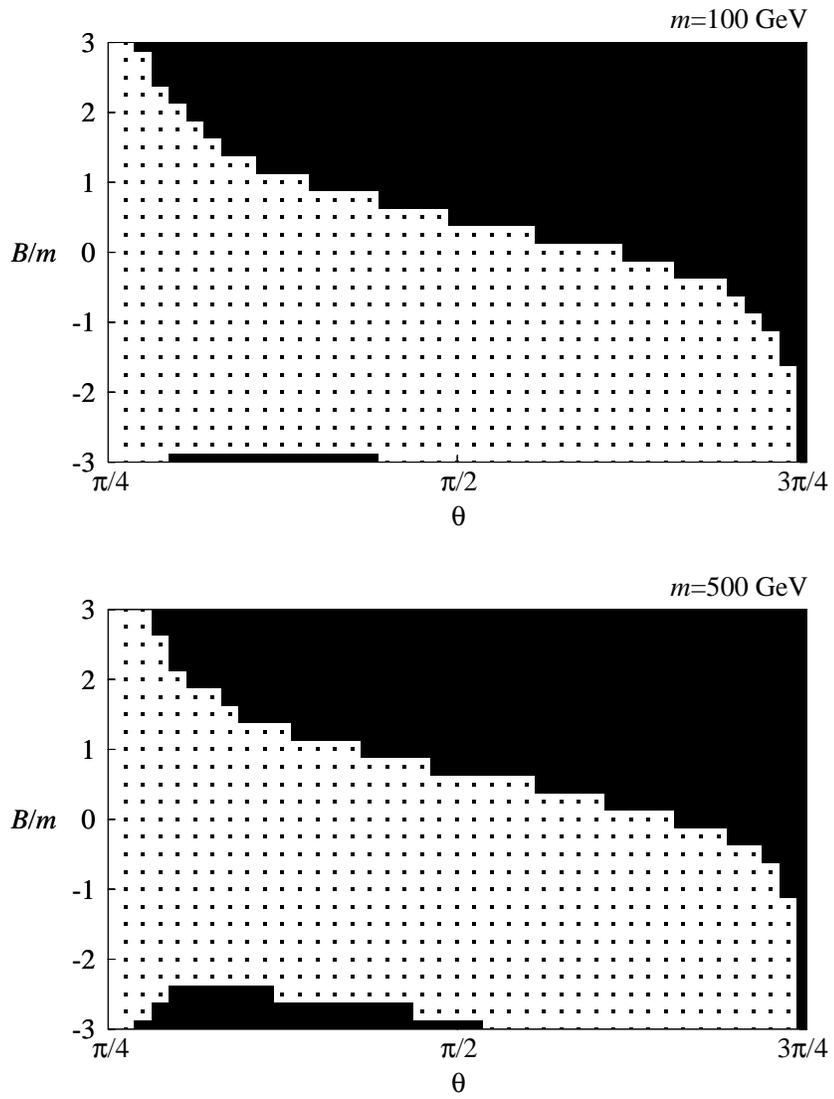

Fig.2
The same as Fig.1a but for a scenario where all the matter fields have modular weight $n_i = -2$ (see subsect. 2.2).

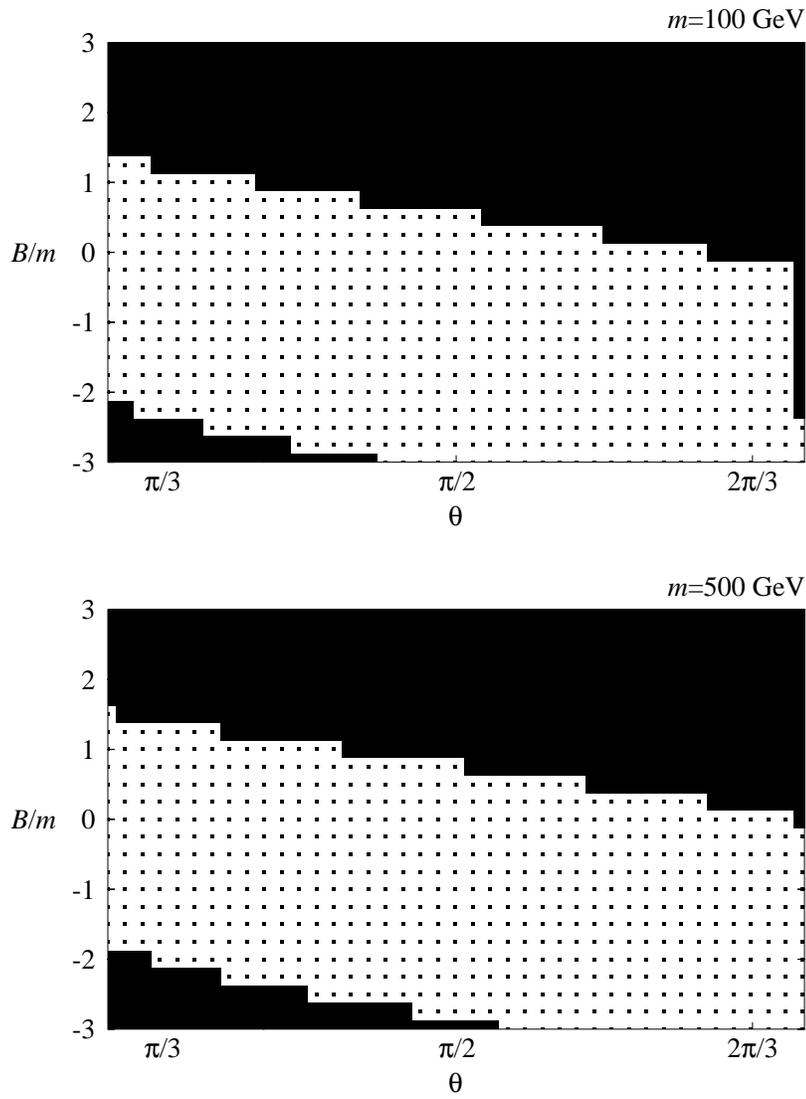

Fig.3
The same as Fig.1a but for the ILR scenario (see subsect. 2.3) in the case $n_{H_1} = -2, n_{H_2} = -3$. $m$ denotes the soft mass of the stop left.

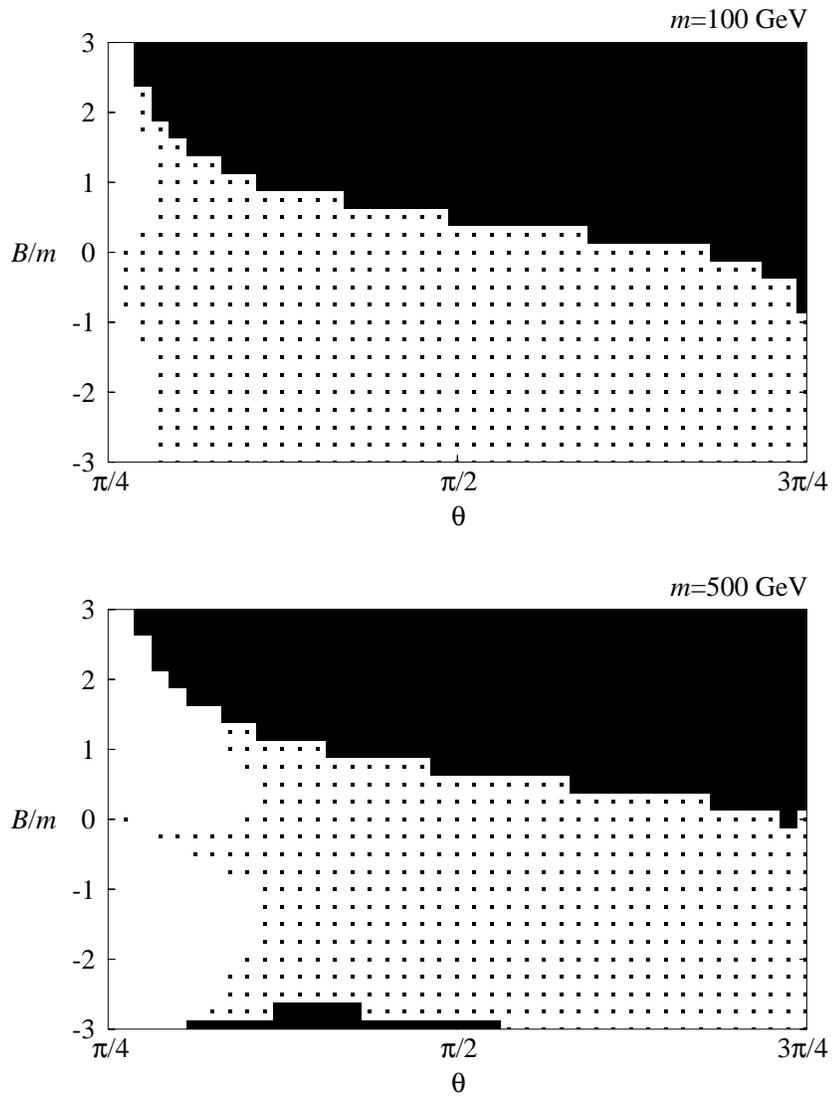

Fig.4a
The same as Fig.1a but for a optimized scenario (see subsect. 2.4) with $n_{H_1} = -1, n_{H_2} = -1$. $m$ denotes the soft mass of the stops.

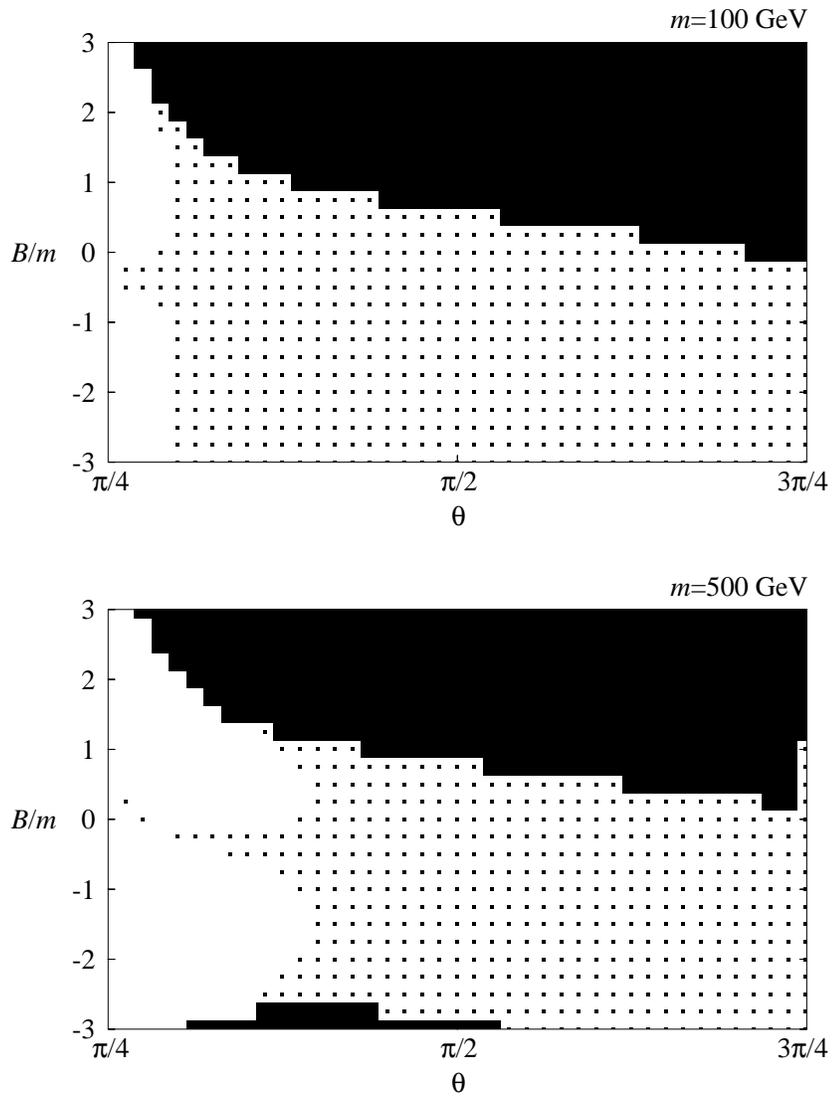

Fig.4b
The same as Fig.4a but assuming non-universal masses for the gauginos in the way explained in the text (see subsect. 2.4).

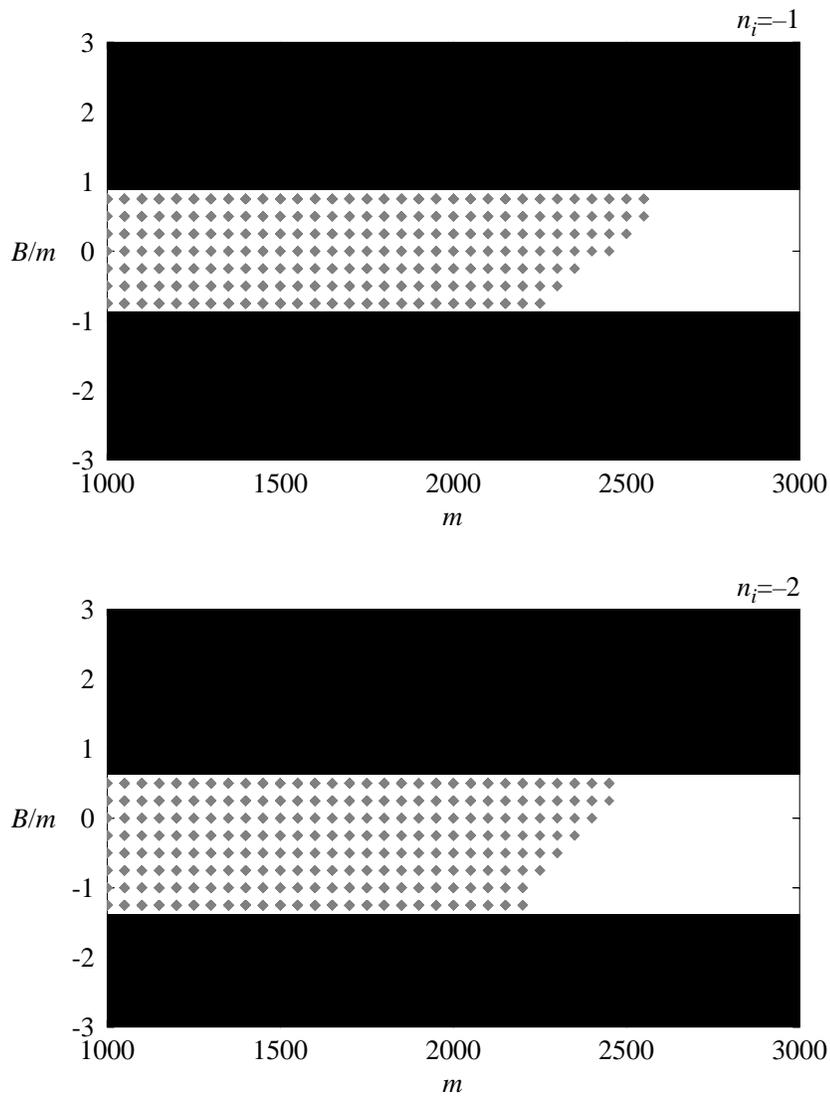

Fig.5
Excluded regions of the parameter space of a weakly coupled string scenario where SUSY is broken by multiple gaugino condensation (see sect. 3). The upper (lower) figure corresponds to the case where all the matter fields have modular weight $n_i = -1$ ($n_i = -2$). The black region is excluded because it is not possible to reproduce the experimental mass of the top, and the diamonds correspond to regions excluded by the experimental lower bounds on supersymmetric particles.

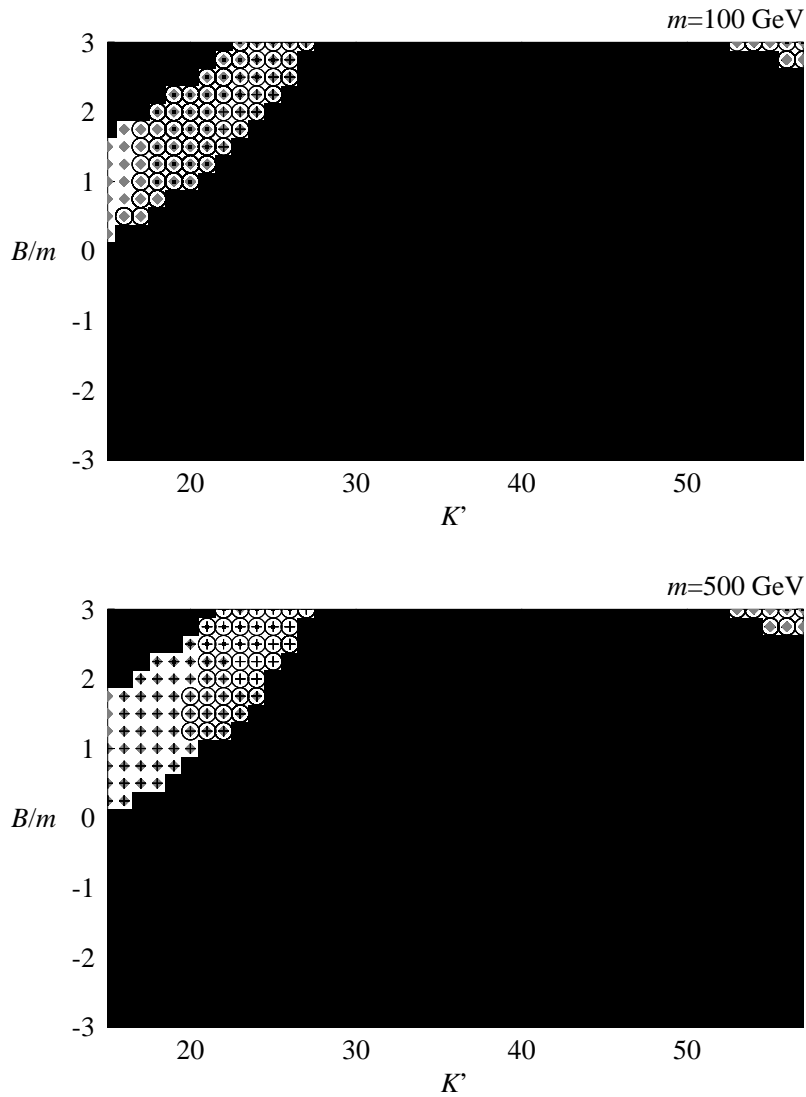

Fig.6
Excluded regions of the parameter space of a scenario where SUSY is broken through a non-perturbative Kähler potential (see sect. 4), for two different values of the soft scalar mass, $m$. The black region is excluded because it is not possible to reproduce the experimental mass of the top; the diamonds correspond to regions excluded by experimental lower bounds on supersymmetric particles; the small squares indicate regions excluded by UFB constraints; the circles indicate regions excluded by CCB constraints, and the crosses indicate regions excluded by negative scalar squared mass eigenvalues.

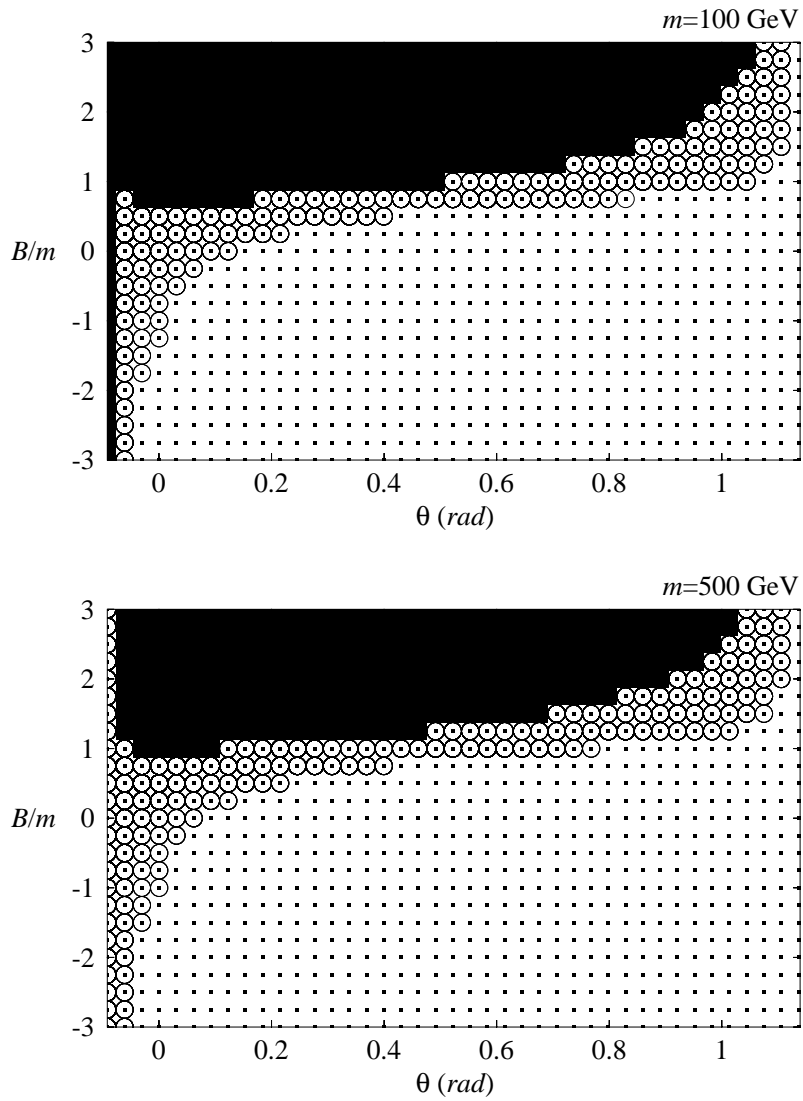

Fig.7a
Excluded regions of the parameter space of a M-theory scenario where $\epsilon = 1$ (see sect. 5), for two different values of the soft scalar mass, $m$. The black region is excluded because it is not possible to reproduce the experimental mass of the top; the small squares indicate regions excluded by UFB constraints; the circles indicate regions excluded by CCB constraints.

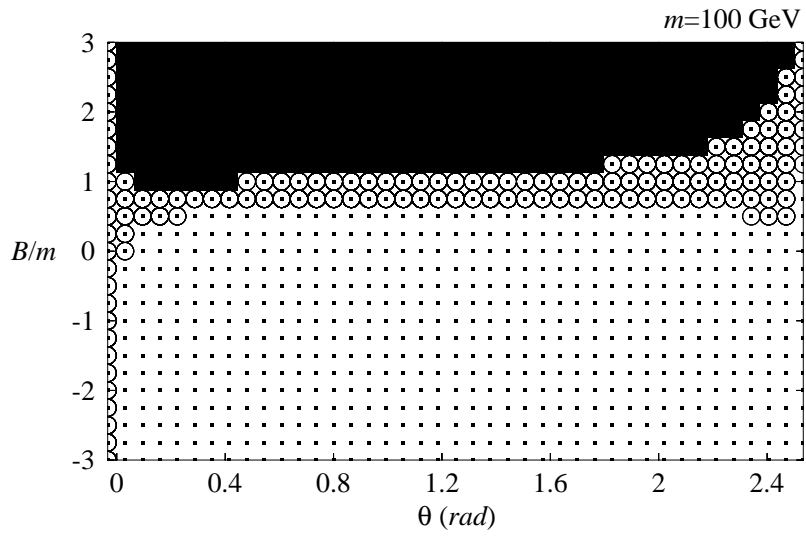

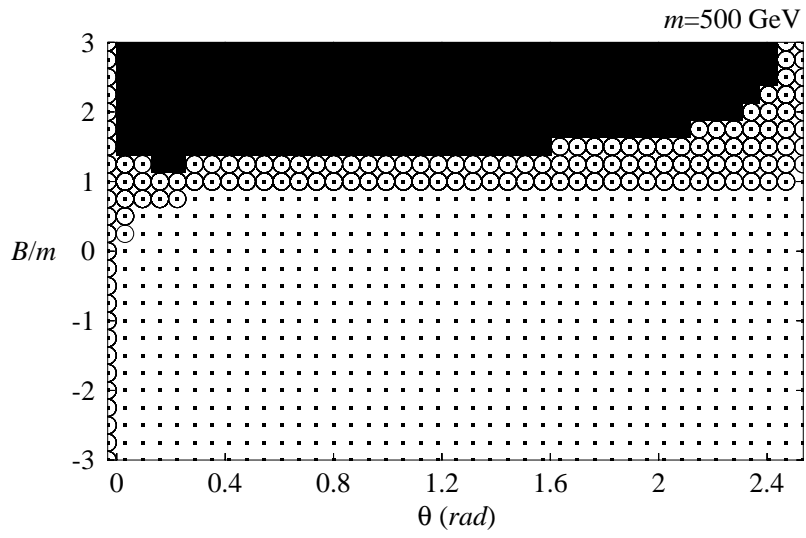

Fig.7b
The same as Fig. 7a but for $\epsilon = 1/3$.